\documentclass[12pt]{article}
\usepackage{amsmath,amsthm,amsfonts,latexsym,amscd,eucal}

\newcommand{\Reali}{{\mathbb{R}}}

\newcommand{\cO}{{\cal O}}
\newcommand{\cC}{{\cal C}}
\newcommand{\cA}{{\cal A}}

\newcommand{\cH}{{\cal H}}

\newcommand{\cL}{{\cal L}}
\newcommand{\cM}{{\cal M}}

\newcommand{\cR}{{\cal R}}

\newcommand{\cV}{{\cal V}}

\def\a{\alpha}
\def\b{\beta}

\def\f{\varphi}

\def\k{\kappa}

\def\L{\Lambda}

\def\o{\omega}
\def\O{\Omega}

\def\r{\rho}
\def\s{\sigma}

\def\x{\xi}

\def\gA{\mathfrak A}

\def\gC{\mathfrak C}

\def\Ppo{{\cal P}_+^\uparrow}

\newtheorem{Thm}{Theorem}[section]
\newtheorem{Cor}[Thm]{Corollary}

\theoremstyle{definition}

\theoremstyle{remark}

\title{The Bisognano-Wichmann Theorem for Charged States\\
and the Conformal Boundary of a Black Hole\footnote{Talk delivered at
the symposium in honor of E.H.
Wichmann ``Mathematical Physics and Quantum Field Theory'', Berkeley, June
1999.}}
\author{Roberto Longo\\
{}\\
\normalsize{Dipartimento di Matematica, Universit\`a di Roma ``Tor
Vergata''}
\\
\normalsize{Via della Ricerca Scientifica, I--00133 Roma, Italy.}
\\
\normalsize{E-mail: longo@mat.uniroma2.it}}
\date{\normalsize{January 12, 2000}}

\begin{document}
\maketitle
\begin{center}
	{\it Dedicated to Eyvind Wichmann on the occation of his seventieth
	birthday}
\end{center}
\bigskip
\bigskip

\noindent
This talk is based on results contained in the following references
\cite{L4,L5,GLRV,L6}. It will concern a study of the incremental
entropy of a quantum black hole, based on Operator Algebras methods.
\section{Case of Rindler spacetime}
Let $\cM=\mathbb R^4$
be the Minkowski spacetime and $\cO\subset \cM \to\cA(\cO)$ be a
net of von Neumann algebras generated by a local Wightman field
on $\cM$. Let $W\subset \cM$ be a wedge region; I may assume
$W=\{x: x_1 > x_0\}$ as any other wedge is a Poincar\'e translated
of this special one. The Bisognano-Wichmann theorem \cite{BW} describes the
modular structure associated with $(\cA(W),\O)$, with $\O$ the vacuum
vector (which is cyclic and separating for $\cA(W)$ by the
Reeh-Schlieder theorem):
\begin{gather}
\Delta^{it}=U(\L_W(-2\pi t)) ,\\
J = U(R)\Theta .
\end{gather}
Here $\Delta$ and $J$ are the Tomita-Takesaki modular operator and
conjugation associated with $(\cA(W),\Omega)$, $\L_W$ is the
one-parameter group of pure Lorentz transformations in the
$x_1$-direction, $U$ is the unitary representation of the Poincar\'e
group, $R$ is the rotation of $\pi$ around the $x_1$-axis and $\Theta$
is the PCT anti-unitary.

In particular, for any fixed $a>0$, the rescaled
pure Lorentz transformations
$\L_W$ give rise to a one parameter automorphism group
$\a_t=\text {Ad}U(\L_W(at))$ of $\cA(W)$
satisfying the KMS thermal equilibrium condition with respect to
$\o|_{\cA(W)}$ at inverse temperature $\b= \frac{2\pi}{a}$, with $\o =
(\O,\cdot\O)$.

As is known the Rindler spacetime may be identified with $W$.
The pure vacuum state $\o$ is thus a (mixed) thermal state
when restricted to the algebra of $W$. Now, if $K= -i\frac{\text
d}{\text{d}t}U(\L_W(t))|_{t=0}$, then $H=aK$ is the proper Hamiltonian
for an uniformly accelerated observer in the $x_1$-direction with
acceleration $a$.
As noticed by Sewell  \cite{S}, the Bisognano-Wichmann theorem
is thus essentially equivalent to the Hawking-Unruh effect in the case
of a Rindler black hole.

Suppose now one adds a short range charge (superselection sector)
localized in the exterior of the black hole: according
to DHR \cite{DHR}, this amounts to consider the net $\cA$ in a different
irreducible localized representation. I may
assume Haag duality to hold and this representation to be realized by a
localized endomorphism $\r$ of the $C^*$-algebra $\gA(\cM)$, where
for an unbounded region $\cR$,  $\gA(\cR)$ denotes the $C^*$-algebra
generated by the $\cA(\cO)$'s with $\cO\subset\cR$ bounded. I may
further choose the charge $\r$ to be
localized in $W$ and to have finite statistics.

Now, as $\r$ is transportable, it turns out that
$\r|_{\gA(W)}$ is normal and I denote by $\r|_{\cA(W)}$ its
weakly continuous extension to $\cA(W)$.
With $\Phi_{\r}$ the left inverse of $\r$ (the unique completely
positive map $\Phi_{\r}:\gA(\cM)\to\gA(\cM)$ with $\Phi_{\r}\cdot\r =
\text{id}$), the state
$\f_{\r}=\o\cdot\Phi_{\r}|_{\cA(W)}$ is clearly invariant under the
automorphism group $\a^{\r}_t=\text{Ad}U_{\r}(\L_W(at))$ of $\cA(W)$,
where $U_{\r}$ is the unitary representation of $\Ppo$ in the
representation $\r$.

The modular structure associated with $\f_{\r}$
is given by the following theorem \cite{L4}.
\begin{Thm}
\item$(i)$ $\a^{\r}$ satisfies the KMS condition with respect to
$\f_{\r}$ at the same inverse temperature $\b=\frac{2\pi}{a}$;
\item $(ii)$ $\log \Delta_{\O,\x_{\r}} = -2\pi K_{\r} +\log d(\r)
= -\b H_{\r} +\log d(\r)$ .
\end{Thm}
Here $K_{\r}$ is the infinitesimal generator of $U_{\r}(\L_W(\cdot))$,
$\xi_{\r}$ is the vector representative of $\f_{\r}$ in the natural
cone of $(\cA(W),\O)$, $d(\r)$ is the DHR statistical dimension of
$\r$ \cite{DHR} and $\Delta_{\O,\x_{\r}}$ is the relative
modular operator.

Now $H_{\r}= aK_{\r}$ is the proper Hamiltonian, in the
representation $\r$, for the uniformly accelerated observer,
so in particular one sees by $(i)$ that the
Bisognano-Wichmann theorem extends with its
Hawking-Unruh effect interpretation in the
charged representation $\r$. This is to be expected inasmuch as
the addition of a single charge should not alter the temperature
of a thermodynamical system.

Point $(ii)$ shows however a new feature: $\log(e^{-\b H_{\r}}\O,\O)$
may be interpreted as the incremental partition function between the states
$\o$ and $\f_{\r}$ on $\cA(W)$. The incremental free energy is then
given by
\[
F(\o|\f_{\r})=-\b^{-1}\log(e^{-\b H_{\r}}\O,\O)\ ,
\]
and indeed it turns out that the thermodynamical relation
$\text{d}F = \text{d}E - T\text{d}S$ holds, namely
\[
F(\o|\f_{\r})=\f_{\r}(H_{\r})-\b^{-1}S(\o|\f{_\r})\ ,
\]
where $S$ is the Araki's relative entropy.

As the statistical dimension takes only integer values by
the DHR theorem \cite{DHR}, one has as a corollary that the values
of the incremental free energy are quantized:
\begin{Cor}\label{cor}
The possible values of the incremental free energy are
\[
 F(\o|\f_{\r})=-\b^{-1}\log n ,\  n=1,2,3,\dots
 \]
 \end{Cor}
The index-statistics theorem \cite{L1} relates $d(\r)$
to the square root of a Jones index \cite{J}; this and other,
see \cite{L4}, leads to the formula
\[
 F(\o|\f_{\r})=-\frac{1}{2}\b^{-1}S_c(\r) ,
 \]
where $S_c(\r)= d(\r)^2$ is the conditional entropy of $\r$.

Notice now that the above discussion is subject to several restrictions.
I have indeed focused on a particular spacetime, the Rindler
wedge $W$, and considered endomorphisms of $\gA(W)$ that are
restriction of endomorphisms of $\gA(\cM)$.

Moreover I have analyzed thermodynamical variables associated with
the exterior of the black hole, without a direct relation to, say, the
entropy of the black hole.

In the following I shall generalize this discussion leading to a
certain clarification of the above aspects.

\section{Case of a globally hyperbolic spacetime}

I shall now consider more realistic black holes spacetimes, namely
globally hyperbolic spacetimes a with bifurcate Killing horizon.

By considering charges localizable on the horizon,
we shall get quantum numbers for the increment of the entropy of
the black hole itself,
rather than of the  outside region.

A main point here is that the restriction of the net $\cA$ of local
observable von Neumann algebras to each of the two horizon components
$\mathfrak h_+$ and $\mathfrak h_-$
gives rise a conformal net on $S^1$, a general fact that is obtained
by applying Wiesbrock's characterization of conformal nets \cite{W},
as already discussed in \cite{L5,GLRV}.

This is analogous to the holography on the anti-de
Sitter spacetime that independently appeared in
the Maldacena-Witten conjecture \cite{Ma,Wi}, proved by Rehren \cite{Re}.
Yet their context differs inasmuch as the anti-de Sitter spacetime is not
globally hyperbolic and the holography there is a peculiarity of that
spacetime,
rather than a general phenomenon.

Finally, I shall deal with general KMS states, besides the
Hartle-Hawking temperature state. In this
context, a non-zero chemical potential \cite{AHKT} can appear, see
\cite{L4}.

The extension of the DHR analysis of superselection sectors
\cite{DHR} to a quantum field theory
on a curved globally hyperbolic spacetime
has been pursued in \cite{GLRV} and I refer to this paper the
necessary background material. I however recall here
the construction of conformal symmetries for the observable
algebras on the horizon of the black hole.

Let $\cV$ be a $d+1$ dimensional
globally hyperbolic spacetime with a bifurcate Killing
horizon. A typical example is given by the Schwarschild-Kruskal
manifold that, by Birkoff theorem, is the only spherically symmetric
solution of the Einstein-Hilbert equation; one
might first focus on this specific example, as the more general
case is treated similarly. I denote by $\mathfrak h_+$ and
$\mathfrak h_-$ the two  codimension 2 submanifolds
that constitute the horizon $\mathfrak h
= \mathfrak h_+\cup \mathfrak h_-$. I assume that the horizon
splits $\cV$ in four connected components, the future, the past and the
``left and right wedges'' that I denote by $\cR$ and
$\cL$ (in the
Minkowski spacetime $\cR= W$ and $\cL=W'$).

Let $\k=\k(\cV)$ be the surface gravity, namely, denoting by
$\chi$ the Killing vector field, the equation on $\mathfrak h$
\begin{equation}
	\nabla g(\chi,\chi)=-2\k\chi\ ,
	\label{k}
\end{equation}
with $g$ the metric tensor,
defines a function $\k$ on $\mathfrak h$, that is actually constant
on $\mathfrak h$ \cite{KW}.
If $\cV$ is the Schwarschild-Kruskal manifold, then
$$
\k(\cV)=\frac{1}{4M},
$$
where $M$ is the mass of the black hole. In this case $\cR$ is
the exterior of the Schwarschild black hole.

Our spacetime is $\cR$ and $\cV$ is to be regarded as a completion of $\cR$.

Let $\cA(\cO)$ be the von Neumann algebra on a Hilbert space $\cH$
of the observables localized in the
bounded diamond $\cO\subset\cR$. I make the assumptions
of Haag duality, properly infinitness of $\cA(\cO)$, Borchers property
B, see \cite{H}.
The Killing flow $\L_t$ of $\cV$
gives rise to a one parameter group of automorphisms $\a$ of the
quasi-local $C^*$-algebra $\gA(\cR)$, since
$\cR\subset\cV$ is a $\L$-invariant region.

I now consider  a locally normal
$\a$-invariant state $\f$ on $\gA(\cR)$,
that restricts to a KMS state at inverse temperature
$\b$ on the horizon algebra, as I will explain. This is clearly
the case of $\f$ is KMS on all $\gA(\cR)$.

The case where $\f$ is the Hartle-Hawking state is of particular
interest; in this case the temperature
\[
\b^{-1} = \frac{\k}{2\pi}
\]
is related to the geometry of the spacetime.

For convenience, I shall assume that the net $\gA$ is already in the
GNS representation of $\f$, hence $\f$ is
represented by a cyclic vector $\x$.
Let's denote by $\cR_a$ the wedge $\cR$
``shifted by'' $a\in\Reali$ along, say, $\mathfrak h_+$ (see \cite{GLRV}).
If $I=(a,b)$ is a bounded interval of $\Reali_+$, I set
$$
\cC(I)=\gA(\cR_a)''\cap\gA(\cR_b)', \quad 0<a<b\ .
$$
One obtains in this way a net of von Neumann algebras on the intervals of
$(0,\infty)$, where the Killing automorphism group $\a$ acts
covariantly by rescaled dilations.

I can now state my assumption:
$\f|_{\gC(0,\infty)}$ is a KMS state with respect
to $\a$ at inverse temperature $\b>0$. Here
$\gC(0,\infty)$ is the $C^*$-algebra generated by all $\cC(a,b), b>a>0$
(for the Schwarzschild spacetime cf. \cite{S}).

It follows that the restriction of the net to the black hole horizon
$\mathfrak h_+$ has many more symmetries than on global spacetime.
\begin{Thm}{\rm (\cite{GLRV}).} The Hilbert space
$\cH_{0}=\overline{\cC(I)\x}$
is independent of the bounded open interval $I$.

The net $\cC$ extends to a conformal net $\mathbb R$ of
von Neumann algebras acting
on $\cH_0$, where the Killing flow corresponds to the rescaled
dilations.
\end{Thm}
This theorem says in particular that one may compactify $\Reali$ to the circle
$S^1$, extend the definition of $\cC(I)$ for all proper intervals
$I\subset S^1$, find a unitary positive energy representation of the
M\"obius group $\text{PSL}(2,\Reali)$ acting covariantly on $\cC$,
so that the rescaled dilation subgroup is the Killing automorphism
group.

If $\x$ is cyclic for $\gC$ on $\cH$, namely $\cH_0 =\cH$
(as is true for the free field in Rindler case, see \cite{GLRV}), then
$\cC$ automatically satisfies Haag duality on $\mathbb R$.
Otherwise one would pass to the dual net $\cC^d$ of $\cC$, which is
is automatically conformal and strongly additive \cite{GLW}.
In the following I assume that $\cC$ is strongly additive.

\subsection{Charges localizable on the horizon}
I now consider an irreducible endomorphism $\r$
of $\gA(\cR)$ with finite dimension that is localizable in
an interval $(a,b)$, $b>a>0$, of $\mathfrak h_+$, namely
$\r$ acts trivially on $\gA(\cR_b)$ and on $\cC(I)$ if $\bar I
\subset (0,a)$, therefore $\r$ restricts to a
localized endomorphism of $\gC(0,\infty)$.

This last requirement is necessary to extend $\r$ to
a normal endomorphism of $\pi_\f(\gC(0,\infty))''$, a result obtained
by conformal methods \cite{L6}.
\begin{Thm} With the above assumptions, if $\r$ is localized
in an interval of $\mathfrak h_+$,
then $d(\r|_{\gC(0,\infty)})$ has a normal extension to the weak closure
$\gC(0,\infty)''$ with dimension $d(\r)$.
\end{Thm}
Let $\s$ be another  irreducible
endomorphism of $\gA$ localized in $(a,b)\subset\mathfrak h_+$ and
denote by $\f_\r$ and $\f_\s$ the thermal states for the
Killing automorphism group in the
representation $\r$. As shown in \cite{L4}, $\f_\r=\f\cdot\Phi_\r$,
where $\Phi_\r$ is the left inverse of $\r$, and similarly for $\s$.

The increment of the free energy between the thermal equilibrium
states $\f_\r$ and $\f_\s$ is expressed as in \cite{L4} by
$$
F(\f_\r|\f_\s)=\f_\r(H_{\r\bar{\s}})-\b^{-1} S(\f_\r|\f_\s)
$$
Here $S$ is the Araki relative entropy and
$H_{\r\bar{\s}}$ is the Hamiltonian on $\cH_0$
corresponding to the charge $\r$ and the charge conjugate to $\bar\s$
localized in $(-\infty, 0)$ as in \cite{L4} (by \cite{GL1}, a transportable
localized endomorphism $\r$ of $\cC$ with finite dimension is M\"obius
covariant). In particular, if $\s$ is the identity
representation, then $H_{\r\bar{\s}}=H_\r$ is the Killing
Hamiltonian in the representation $\r$.

By the analysis in \cite{L6} one can write formulas
in the case of two different KMS states $\f_\r$ and $\f_\s$.
In particular
\begin{equation}
F(\f_\s|\f_\r)=\frac{1}{2}\b^{-1}(S_c(\s)-S_c(\r)) + \mu(\f_\s|\f_\r),
\end{equation}
where $\mu(\f_\s|\f_\r)$ is the chemical potential, cf. \cite{AHKT}.

The above formula gives a canonical splitting for the incremental
free energy. The first term is symmetric under charge conjugation,
the second term is anti-symmetric. The quantization of the possible values
of the symmetric part goes through
similarly as in Corollary \ref{cor}.

\section{Quantum index theorem}
The above discussion is part of a general aim concerning a quantum
index theorem. This point of view is discussed in \cite{L6}.


{\footnotesize \begin{thebibliography}{7}

\bibitem{AHKT} H. Araki, R. Haag, D. Kastler, M. Takesaki:
{\it Extension of KMS states and chemical potential}, Commun.
Math. Phys. {\bf 53}, 97-134 (1977).

\bibitem{BW}  J. Bisognano, E. Wichmann: {\it On the duality condition
for a Hermitean scalar field}, J. Math. Phys. {\bf 16},  985 (1975).


\bibitem{DHR}   S. Doplicher, R. Haag, J.E. Roberts: {\it Local
observables and particle statistics I}, Commun. Math. Phys.
{\bf 23},  199-230 (1971), II Commun. Math. Phys.
{\bf 35}, 49-85 (1974).

\bibitem{GL1} D. Guido, R. Longo: {\it Relativistic invariance and
charge conjugation in quantum field theory},
Commun. Math. Phys. {\bf 148}, 521 (1992).

\bibitem{GLW}  D. Guido, R. Longo, H.-W. Wiesbrock: {\it Extension of
conformal nets and superselection structures}, Commun. Math. Phys.
{\bf 192},  217-244 (1998).

\bibitem{GLRV} D. Guido, R. Longo, J.E. Roberts, R. Verch:
{\it Charged sectors, spin and statistics in quantum field theory
on curved spacetimes}, Rev. Math. Phys. (to appear).

\bibitem{H} R. Haag ``Local Quantum Physics", Springer-Verlag
(1996).

\bibitem{HL}  P.D. Hislop, R. Longo: {\it Modular structure of the local
algebras associated with the free massless scalar field theory},
Commun. Math. Phys. {\bf 84}, 71 (1982).


\bibitem{J}V. F. R. Jones: {\it Index for subfactors},
Invent. Math. {\bf 72} (1983) 1-25.

\bibitem{L1} R. Longo: {\it Index of subfactors and statistics of
quantum fields.  I \& II}, Commun.  Math.  Phys.  {\bf 126}, 217--247
(1989) \& Commun. Math. Phys. {\bf 130},  285-309 (1990).

\bibitem{L4} R. Longo: {\it An analogue of the Kac-Wakimoto
formula and black hole conditional entropy}, Commun. Math. Phys.
{\bf 186} (1997), 451-479.

\bibitem{L5} R. Longo: Abstracts for the talks at the Oberwolfach
meetings on ``$C^*$-algebras'', February 1998, and ``Noncommutative
geometry'', August 1998.

\bibitem{L6} R. Longo: {\it Notes for a quantum index theorem},
manuscript, 1999.

\bibitem{Ma} J. Maldacena,  {\it The large N limit of superconformal
field theories and supergravity},  Adv. Theor. Math. Phys. {\bf 2}
(1998), 231-252.

 \bibitem{Re} H.K. Rehren: {\it Algebraic holography},  Preprint,
 May 1999.

\bibitem{S} G.L. Sewell: {\it Quantum fields on manifolds:
PCT and gravitationally induced thermal states}, Ann. Phys.
{\bf 141} (1982), 201.

\bibitem{KW} R.M. Wald, ``Quantum Field Theory in Curved Spacetime
and Black Hole Thermodynamics'', University of Chicago
Press, 1994.

\bibitem{W} H.-W. Wiesbrock: {\it Conformal quantum field theory and
half-sided modular inclusions of von Neumann algebras}, Commun. Math. Phys.
{\bf 158} (1993), 537.

\bibitem{Wi} E. Witten: {\it Anti de Sitter space and holography},
Adv. Theor. Math. Phys. {\bf 2}
(1998), 253-291.

\end{thebibliography}}
\end{document}